\documentstyle[multicol,aps,psfig]{revtex}

\newcommand{\beq}{\begin{equation}}
\newcommand{\eeq}{\end{equation}}
\newcommand{\bdis}{\begin{displaymath}}
\newcommand{\edis}{\end{displaymath}}
\newcommand{\bea}{\begin{eqnarray}}
\newcommand{\eea}{\end{eqnarray}}
\newcommand{\barr}{\begin{array}}
\newcommand{\earr}{\end{array}}
\begin{document}

\title{Kinetic roughening with anisotropic growth rules}
\author{Raffaele Cafiero$^1$}

\address{
$^1$L.P.M.M.H. Ecole Sup\'erieure de Physique et de Chimie Industrielles,
10, rue Vauquelin, 75231 Paris CEDEX 05 France}
\maketitle
\date{\today}
\begin{abstract}
Inspired by the chemical etching processes, where experiments
show that growth rates depending on the local environment might play a
fundamental role in determining the properties of the etched surfaces,
we study here a model for kinetic roughening which includes explicitly
an anisotropic effect in the growth rules.  Our model introduces a
dependence of the growth rules on the local environment conditions,
i.e. on the local curvature of the surface. Variables with different
local curvatures of the surface, in fact, present different quenched
disorder and a parameter $p$ (which could represent different experimental
conditions) is introduced to account for different time scales for the
different classes of variables. 
We show that the introduction of this {\em time scale separation} 
in the model leads to a cross-over effect on the roughness properties. 
This effect could explain the scattering in the experimental measurements 
available in the literature. 
The interplay between  anisotropy and the cross-over effect 
and the dependence of critical properties on parameter $p$ is 
investigated as well as the relationship with the known 
universality classes. 
\end{abstract}
\smallskip


\begin{multicols} {2}

\section{Introduction}

In the last few years the study of physical 
phenomena characterized by a degree of self-organization \cite{btw}, 
has attracted a lot of interest. These models are usually cellular 
automata models defined on a discretized lattice, with a growth rule 
that can be either stochastic, when the inhomogeneities in the system 
change with a time scale smaller than the characteristic time scale of 
the dynamical evolution (noise), or deterministic with a quenched 
disorder which accounts for the effect of inhomogeneities inside a solid 
medium. Both kind of dynamical rules are characterized by an evolution 
towards an attractive fixed point in which scale free fluctuations in 
time and space are present \cite{soc}.

The problem of kinetic roughening belongs to this class of models. 
It received recently an increasing interest in relation with 
non-equilibrium growth models \cite{zhang} 
and in view of its practical applications: Chemical Vapor 
Deposition (CVD) \cite{CVD} and electro-chemical deposition 
\cite{electro} are just two examples. 

In this perspective, one is interested 
in identifying the dynamic universality classes of kinetic 
roughening processes and  
several models has been defined starting 
from the models falling in the universality class of the 
Kardar-Parisi-Zhang (KPZ) equation \cite{kpz}. This equation 
describes the properties of an interface $h(x,t)$ driven by a stochastic 
noise and gives a roughness exponent $\chi=0.5$. Other models 
are more suitable to describe the propagation of interfaces 
in random media, i.e. with a 
quenched disorder. These models are driven by an extremal dynamics.
In this class fall the so-called Sneppen model \cite{snep} 
(in \cite{snep} referred as model B) and the pinning model by directed 
percolation \cite{tang}, which predict a roughness exponent equal 
to $\chi=0.63$. These models produce self-affine surfaces. 
Recently, a model has been introduced to describe some 
etching experiments, which leads to
 the formation of self-similar (fractal) structures, and which has 
been shown to fall in the percolation universality class \cite{sapoval}. 
Many experiments on surface roughening \cite{burn,fluid,horvath}, however, 
as well as experiments on chemical etching \cite{etchlett} 
produce self-affine surfaces instead of self-similar ones. In this paper, 
we will focus on kinetic roughening phenomena 
leading to {\em self-affine} ($\chi < 1$) surfaces.

We recall that the roughness exponent is defined by the ensemble 
averaged width of the interface as 
$W(l,t)= <(h(x,t)-<h(x,t)>)^2>^{1/2} \sim l^{\chi} f(t^{1/z}/l)$
where $z$ is the so-called dynamical exponent, the angular brackets
denote the average over all segments of the interface of length $l$ 
and over all different realizations. $f$ is a scaling
function such that $f(y) \sim y^{\chi}$  for $y<<1$ and $f(y)=cost.$
for $y>>1$. The exponent $\beta=\chi/z$ describes the transient
roughening, during which the surface evolves from the initial 
condition toward the final self-affine structure.

In spite of the strong universality exhibited by the KPZ 
and Sneppen models, in many 
experimental studies one measures
 values of $\chi$ which are above the ones predicted 
by both the KPZ and Sneppen universality classes. 
To give some examples, we remember the experimental 
studies of paper burning for which one gets 
$\chi=0.70\pm0.03$ \cite{burn}, or the 
propagation of a forced fluid front in a porous 
medium, which exhibits a roughness exponent $\chi=0.73\pm0.03$ 
\cite{fluid} and $\chi=0.88\pm0.08$ \cite{horvath}. 

In this paper we propose a generalized model for kinetic roughening 
characterized by anisotropic growth rules and, as a consequence, 
separated time scales for the dynamics. The existence of this time-scale 
separation induces a cross-over effect in the roughness properties, 
which could  erroneously appear as a genuine non-universal critical 
behavior, and could give an explanation for the above cited scattered 
experimental results.

Some results presented here have been already briefly
reported on in a letter \cite{etchlett}. In this long paper 
we give a detailed, complete description of our previous work. 
Moreover, we present a set of new numerical results, 
which allow us to reach different and better founded conclusions 
with respect to \cite{etchlett}. 

The idea underlying the model is that some experimental 
parameters can introduce a characteristic scale in the system, 
separating different scaling behaviors.  
In particular we consider a
model which includes explicitly an anisotropy factor,
say a growth rule dependent on the local environment of the 
growing site. The model thus presents a complex interplay
between a global equilibrium and the conditions of a 
local dynamics. 
This choice is motivated by the observation
of roughening phenomena occurring in etching processes
which represent an important tool either in academic research 
or in device technology. Their importance is related
to the preparation of single-crystal samples 
of desired dimensions, shapes and orientations. 
Etching is usually applied to obtain
desired mesas and grooves in semiconductors wafers and 
multilayers \cite{yablo}.

In the same field, although in a different 
context, etching processes are
 used to produce textured optical sheets, which allow to exploit the 
light  trapping by total internal reflection to increase the effective
 absorption in the indirect-gap semiconductors crystalline silicon. 
Light trapping, originally suggested to 
increase the response speed of
silicon photo-diodes while maintaining high quantum efficiency
in the near-infrared, was later indicated as 
an important benefit for solar cells \cite{yablo}.

The general suffix {\em etching} indicates the ensemble of
operations which involve the removal of materials by expending
 energy either by mechanical, thermal or 
chemical means. In ref. \cite{etchlett} the authors 
focused their attention on chemical etching processes 
as a reference point to formulate the model.
One of the most important properties of these processes is
represented by the intrinsic anisotropy \cite{seidel,elwen} 
of the etch rates.
For instance in samples of crystalline silicon etched 
in solutions of aqueous 
potassium-hydroxide (K-OH) with isopropil 
alcohol (IPA), depending on the concentration of the etchant 
and the temperature, the $(111)$ direction etches slower than the 
others by a factor which can be of order $100$ or more \cite{allongue}. 
The degree of anisotropy affects the properties of the surface, 
which turns out to be rough with an apparently non-universal roughness 
exponent.

Although the definition of the model is very general
we will briefly consider the chemistry of the etching process
in order to exhibit a physical framework that allows to
understand the meaning of the definitions and their interpretation.

The disorder in the etching process is related to the impurities in 
the lattice. Such impurities, e.g. vacant atoms, reduce the binding 
energy of atoms nearby the vacancy. By assigning to each site (atom) 
of our lattice a random number $x_i$ we assume that a distribution of 
vacancies, or other kinds of impurities, is present in the system, and 
this induces fluctuations in the binding energy of atoms due to this 
disorder. If we assume to be in a condition of slow dynamics, that is to 
say the driving field (which in our case in represented by the 
concentration of etchant) tends to zero \cite{slow}, we can look at the 
etching as an extremal process, 
where the etchant dissolves the atom with the smallest binding 
energy. This is correct for low etchant concentrations
and corresponds actually to the situation experimentally 
more interesting, in which rough surfaces are produced.

In order to reproduce the experimental conditions (type of etchant, 
concentration, temperature) a microscopic model 
for the physical process should contain some tunable parameters (at least one).
In our model, the anisotropy is introduced by a phenomenological tunable 
parameter, $p$, which distinguishes sites with a different local environment.

The introduction of the parameter $p$ defines a characteristic scale in the 
problem. As a result the critical properties of the model are characterized 
by a continuous crossover between two universality classes corresponding 
to the roughness exponents $\chi=1$ and $\chi=0.63$ (Sneppen models $A$ and $B$
respectively \cite{snep}).  In particular one can define a parameter $r=\frac{p}{1-p}$
which measures the time-scale separation between the dynamics of the different classes
of sites. For lengths $l > l^* \propto \frac{1/2-p}{p}$ one observes a behavior characteristic 
of the Sneppen model $B$ universality class ($\chi=0.63$), while 
for $l < l^* \propto \frac{1/2-p}{p}$ one observes a behavior characteristic 
of the Sneppen model $A$ universality class.
The existence of this crossover is difficult to detect directly 
on the plot for the scaling of the surface width  $W^2(l)$ (especially
for finite sets of data) and it becomes evident looking at the power spectra.
This explain why large-scale experiments could give the impression
of non-universality in the critical properties of rough surfaces.

\begin{figure}[h]
\centerline{
       \psfig{figure=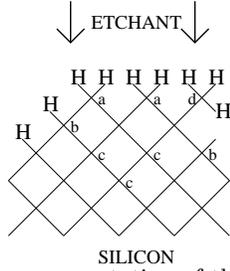,width=3cm,angle=-90}}
\protect\caption{Schematic representation of the 
crystalline silicon lattice as a square lattice.}
\label{fig1}
\end{figure}

Let us look at the meaning of $p$ in the case of etching. 
If we represent the crystalline lattice of the silicon on a two-dimensional 
plane we can imagine a square lattice where (see Fig.(\ref{fig1})) the 
atoms can be found in each of the four positions marked in figure by the letters 
$a-d$. The four positions correspond to different oxidation states: from the situation 
$(c)$ (oxidation number $0$) which occurs only in the bulk, to the situation $(d)$ 
(oxidation number $-3$).  Note that all the surface atoms are 
passivated by hydrogen atoms. The atoms in the positions $(a)$ and $(b)$, 
corresponding respectively to the oxidation numbers
$-2$ and $-1$ (two and one heteropolar bonds, i.e. Si-H bonds), 
play an important role in explaining, at least from a heuristic point 
of view, the origin of the anisotropy in the etched rates \cite{elwen}.
The parameter $p$ quantifies the ratio of the etch rates between
the sites in the positions $(a)$ and $(b)$. 
The basic idea is that in the $(111)-$plane of silicon, there is 
only one heteropolar bond per silicon atom. 
Therefore there are three bonds to break for dissolution, 
while other planes (except the $(110)$) have more than one
heteropolar bonds and accordingly a smaller number of
bonds must be broken.

The paper is organized as follows. In section II we describe in detail 
the model and the setup of numerical simulations. In section III we 
present and discuss the numerical results in relation with the Sneppen 
model A and B universality classes. 
In section IV a discussion of the results and 
some conclusions are drawn together with a planning of future 
researches.

\section{The model}

We give now a detailed definition of the model.
The model is defined on a square $2D$ lattice tilted at $45^{\circ}$ 
(see Fig. \ref{fig1}).
We consider a $1+1$ dimensional interface $h(x)=h(x,t)$ with 
$x=1,2,...,L$, where $L$ is the linear extension of the interface in the 
$x$ direction. The initial condition for the dynamical evolution of this 
interface is given by:
$h(2x,0)=1 \,\, \forall x \in [1, L/2 ]$ and 
$h(2x-1,0)=0 \,\, \forall x \in [1, L/2 ]\,\, ,$
in order to have both classes of variables (lattice planes) 
participating to the dynamics from the beginning, 
but different initial conditions do not change the properties 
of the model.
The interface, which satisfies locally the conditions \cite{kim}
$$\vert h(x,t)+1-h(x-1,t) \vert \le 1,$$ 
\begin{equation}
\vert h(x,t)+1-h(x+1,t) \vert \le 1,
\label{kk}
\end{equation}
contains two classes of random variables that correspond to two separate classes of 
sites. 
The sites $(M)$ for which it holds $\nabla^2 h >0$ (called minimum sites)
which are, microscopically, the atoms with two heteropolar bonds, and
the sites on a slope (slope $(S)$ sites) for which one has $\nabla^2 h = 0$. 
These last sites correspond microscopically to atoms with one heteropolar bond.
To each class of sites is assigned a class of Gaussian distributed 
uncorrelated random variables which mimic the disorder, and 
represents physically, for the case of etching, the binding energy of atoms:
\beq
\eta (x,h) \in \left\{ 
\begin{array}{lc}
\left[ 0:0.5 \right] &  \mbox{if x is such that} \,\,  \nabla^2 h =0\\
\left(0.5:1 \right] & \mbox{if x is such that} \,\, \nabla^2 h > 0. 
\end{array}
\right.
\eeq
The sites  with  $\nabla^2 h  < 0$, for which all the chemical bonds 
are homopolar, i.e. Si-Si, do not take part to the dynamics and
they have assigned a zero value of the random variable. 
Periodic boundary conditions are assumed along the $x$ direction.

\begin{figure}[h]
\centerline{
        \psfig{figure=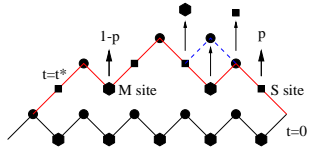,width=4cm,angle=-90}}
\protect\caption{
Schematic representation of interface dynamics.}
\label{fig1bis}
\end{figure}

The system evolves by updating
the site $i^*$ with the 
{\bf largest} r.v. in one of the two classes of 
sites chosen, at its turn, with a probability $p$. 
One thus updates with probability $p$ a site 
$(S)$ and with probability 
$1-p$ a site $(M)$ according to the rules (see Fig. (\ref{fig1bis})):
\begin{itemize}
\item[(1)] $h(i^*, t+1)=h(i^*,t)+2$, $\,\,$ $\eta(i^*,h(i^*,t+1),t+1)=0$;
\item[(2)] Updating all the sites necessary to make satisfied the
conditions \ref{kk} (this phase is assumed to be instantaneous 
with respect to extremal dynamics);
\item[(3)] Updating of the random variables for the sites which 
changed their class of belonging. In particular
$\eta (x,h,t+1) = 1/2 * RAN$ if $\eta(x,h,t)=0$ and
$\eta (x,h,t+1)= \eta(x,h,t)+1/2$ if $\eta(x,h,t) \ne 0$,
where $RAN$ is a random value between $0$ and $1$;

\item[(4)] Updating of the random variables of 
the sites which have changed their height but which 
did not change their class  of belonging, (sites S only): 
$\eta (x,h,t+1) = 1/2 * RAN$.

\end{itemize}

The parameter $p$ can vary in the range 
$[0:1/2]$. If we define $t_S$ as the characteristic time scale
 for $S$ variables and $t_M$ the characteristic time scale 
for $M$ variables one has:
\beq
 r=\frac{t_M}{t_S}=\frac{p}{1-p}
\label{scsep}
\eeq
The growth of the interface, in the etching process, represent 
the invasion of the etchants into the silicon wafer. 
>From this point of view the updating of the sites $(S)$ mimics 
the etching of the $(111)$ planes and the updating of the $(M)$ sites 
the etching in the $(100)$ direction.
For $p=1/2$ all the sites which take part to the dynamics are
 equivalent and there is no anisotropy ($r=1$), whereas the case $p=0$
corresponds to the maximal anisotropy in which
$v_{(111)}/ v_{(100)} =r=0$, where $v_{(111)}$ and $v_{(100)}$
are the etch rates in the corresponding directions.

Our model can be viewed as a variation of the Sneppen model for quenched
surface growth, where two important elements are added: 1) The anisotropy 
in the distribution of the quenched random field, depending on the local
characteristics of the growing surface: 2) a time scale separation for the
dynamical evolution of the two classes of variables ($S$ sites and $M$ sites),
which is tuned by the parameter $p$.

\section{Numerical results}

We have studied this cellular automata by numerical 
simulations in order to analyze its dynamical roughening 
properties. The sizes we have chosen for the 
numerical simulations range from $L=2048$ to $L=8192$.
For each value of $p$(we have considered $p=0.0,0.02,0.2,0.5$) , 
$10^2$ simulations lasting $10^7$ time  steps have been performed
and we have computed the growth exponent $\beta$, which rules the time evolution 
of the width $W(t)$ of the surface ($W(t)\sim t^{\beta}$)
before the stationary state is reached, and the roughness exponent 
$\chi$, which gives the scaling of the width of the surface ($W(l)\sim l^{\chi}$), 
in the stationary state. The stationary state is called self-organized in that
it is reached spontaneously by the system independently of the 
initial conditions. This self-organization is confirmed by an 
analysis of the temporal evolution of the distribution of quenched variables
(the histogram $\Phi_t(\eta)$). To characterize temporal correlations
in the dynamics and check that the asymptotic state is critical, we
studied the distribution of the avalanches in the asymptotic state.
As an independent check about the universality of the 
roughness properties of the model, we have studied numerically the 
power spectrum $S(k)$ of the height profile.

\begin{figure}[h]
\centerline{
        \psfig{figure=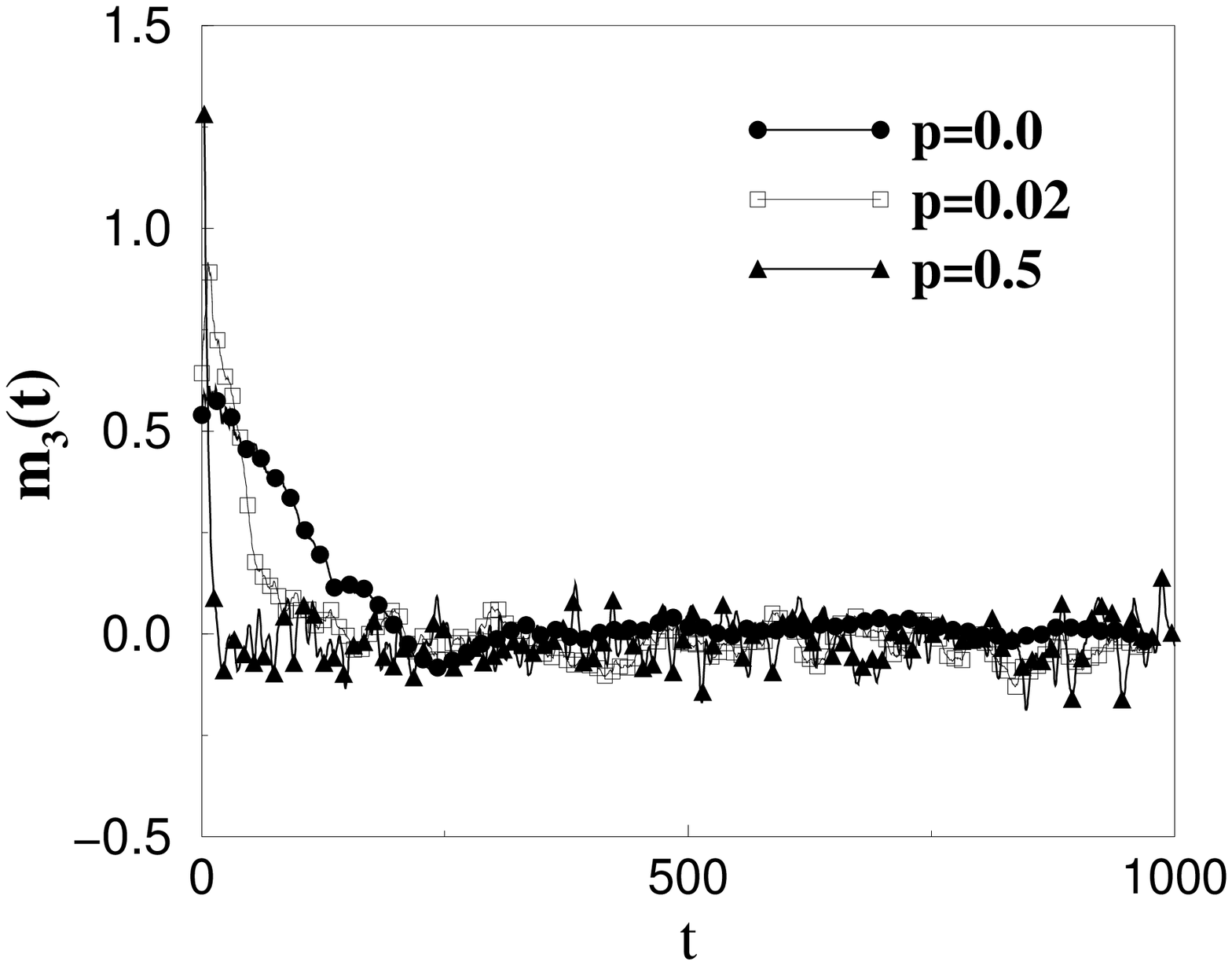,height=5.5cm,angle=0}}
\protect\caption{Time evolution (adimensional units) of the moment $m_3$ 
 (adimensional units) of the growing interface, normalized by
the second moment, for $p=0.0,0.02,0.5$ (skewness). 
One sees that asymptotically $m_3$ vanishes for all values of $p$.}
\label{mom3}
\end{figure}

\begin{figure}[h]
\centerline{
        \psfig{figure=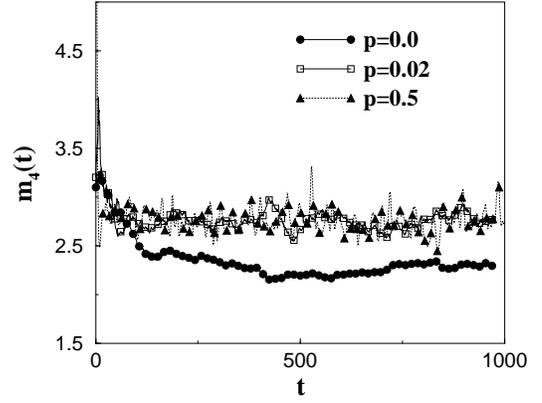,height=5.5cm,angle=0}}
\protect\caption{Time evolution (adimensional units) of the moment $m_4$ 
 (adimensional units) of the growing interface, normalized by
the second moment, for $p=0.0,0.02,0.5$. 
One sees that asymptotically $m_4$ tends to different constant 
values for the different values of $p$.}
\label{mom4}
\end{figure}

\begin{figure}[h]
\centerline{
        \psfig{figure=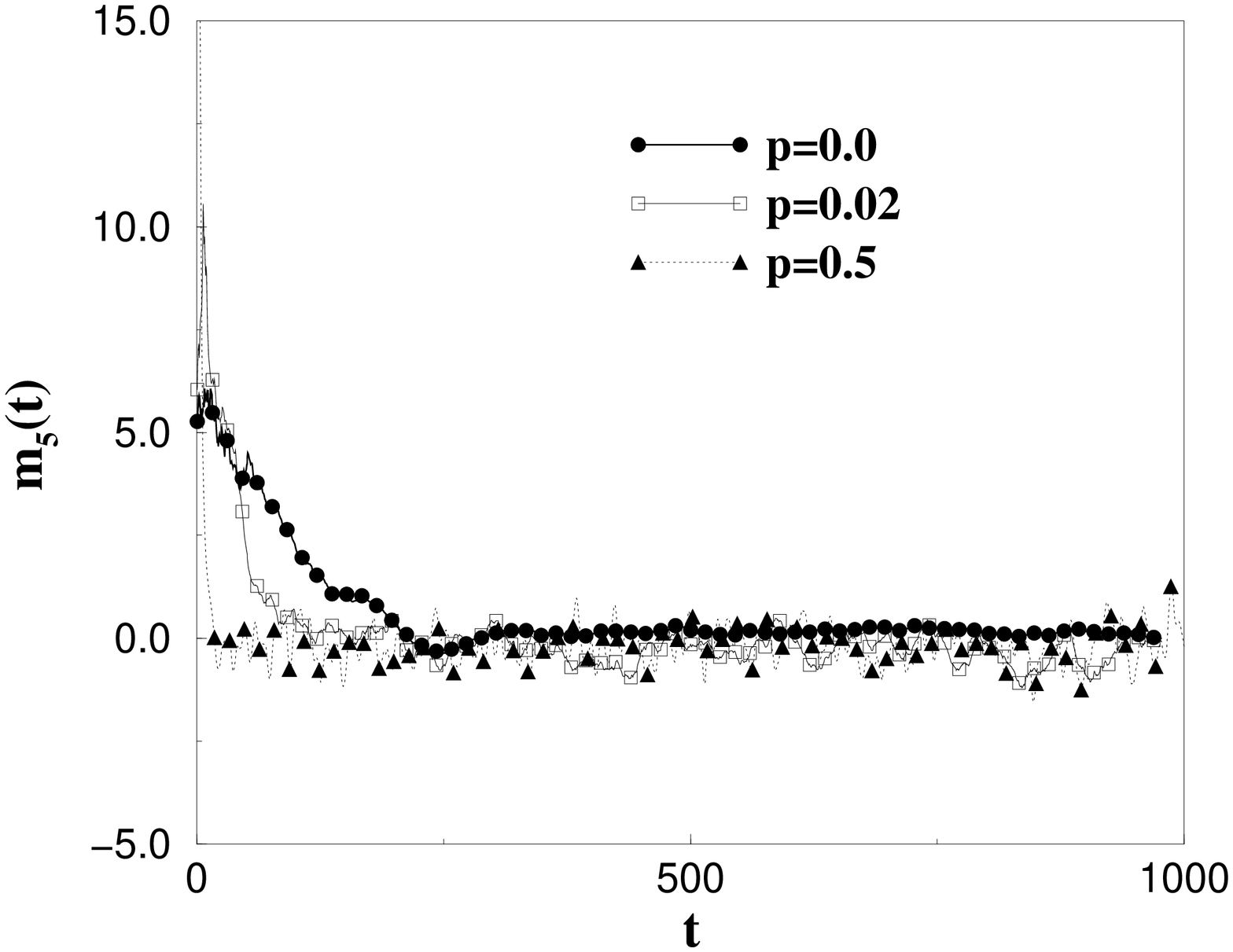,height=5.5cm,angle=-90}}
\protect\caption{Time evolution (adimensional units) of the moment $m_5$ 
(adimensional units)  of the growing interface, normalized by
the second moment, for $p=0.0,0.02,0.5$. 
One sees that asymptotically $m_5$ vanishes for all values of $p$.}
\label{mom5}
\end{figure}

To ensure that the system is in the stationary state, 
we studied the behavior of the $n$-th moments (for $n=3,4,5$) 
of $h(x,t)$ normalized to second momentum:
\begin{equation}
m_n(t)=\frac{\langle{\sum_i(h(i,t)-\bar{h}(t))^n}\rangle}
{\left(\langle{\sum_i(h(i,t)-
\bar{h}(t))^2}\rangle\right)^{\frac{n}{2}}},
\label{momenti}
\end{equation}
where $\bar{h}(t)$ is the mean surface height at time $t$.
We get that, after a transient, all the odd moments
vanish and the even ones tend to constant values 
(see Fig.s \ref{mom3}-\ref{mom5}). 
In particular the condition for the skewness $m_3=0$ (Fig. \ref{mom3}), 
which characterizes the stationary critical state\cite{snep}, 
is realized after about $10^3$ time steps per site, 
independent of the value of $p$. 
These results imply that the higher moments of the variable 
$h(x,t)-\langle{h}\rangle$ scale in a trivial way (they are 
powers of the second moment), and after the 
transient the probability distribution of the variable $h(x,t)$, which 
can be viewed as a random variable, is Gaussian. The amplitude of the 
normalized even moments $m_{2n}$, in the asymptotic stationary state, 
characterizes the roughness properties of the interface.

\begin{figure}[hb]
\centerline{
            \psfig{figure=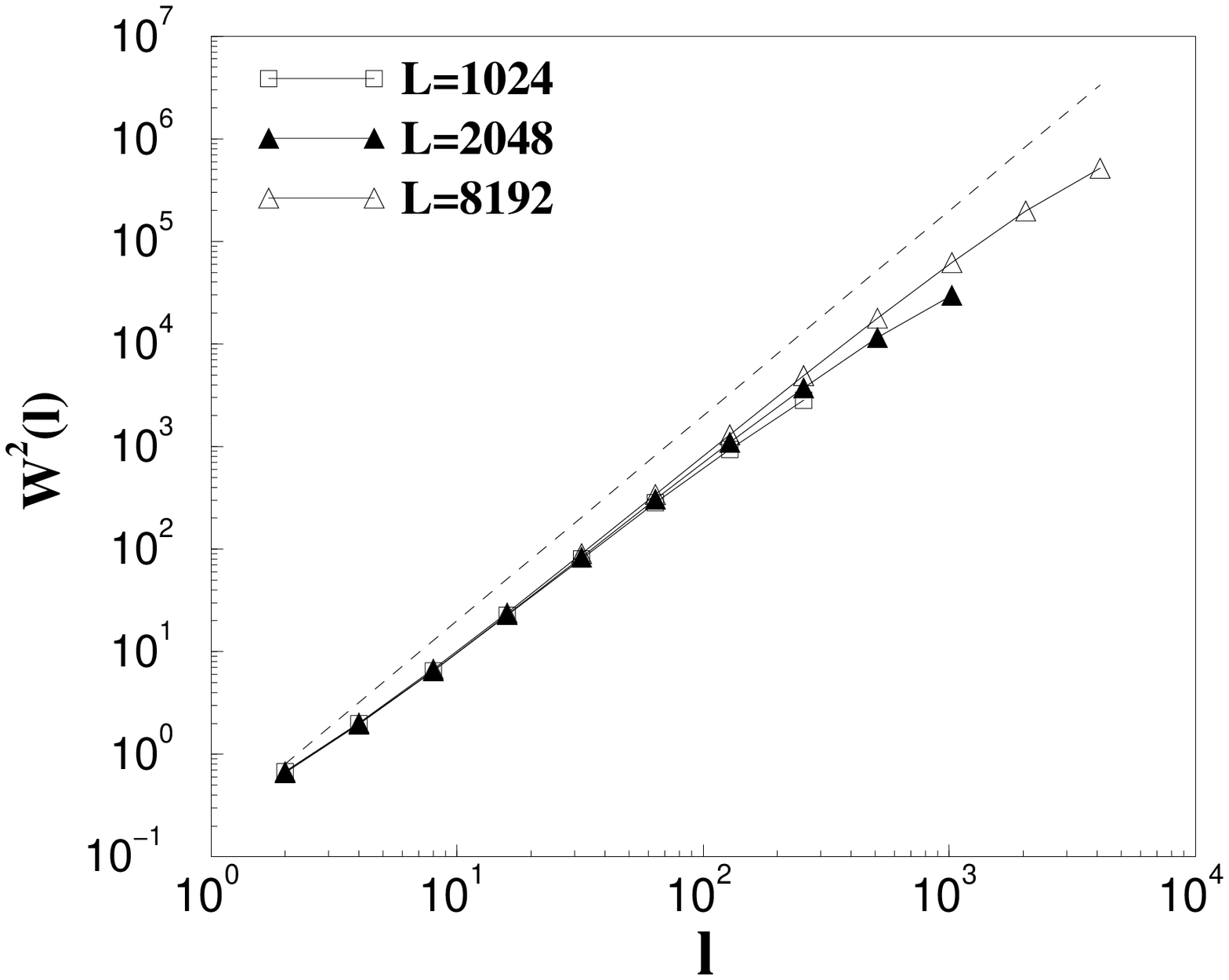,height=5.5cm,angle=0}}
\protect\caption{$W^2(l)$ vs. $l$  (both $W^2(l)$ and 
$l$ are expressed in adimensional units) for $p=0.0$ and different 
system sizes. The lower fit
(dot-dashed line) corresponds to a size $L=512$, 
giving and exponent $\chi=0.88(2)$, 
while the upper fit (dashed line) corresponds 
to a size $L=8192$ and gives an exponent $\chi=0.96(2)$.
In this case we expect that the exponent converges to
$\chi=1$ in the limit $L \rightarrow \infty$.}
\label{finite}
\end{figure}

\begin{figure}[hb]
\centerline{
            \psfig{figure=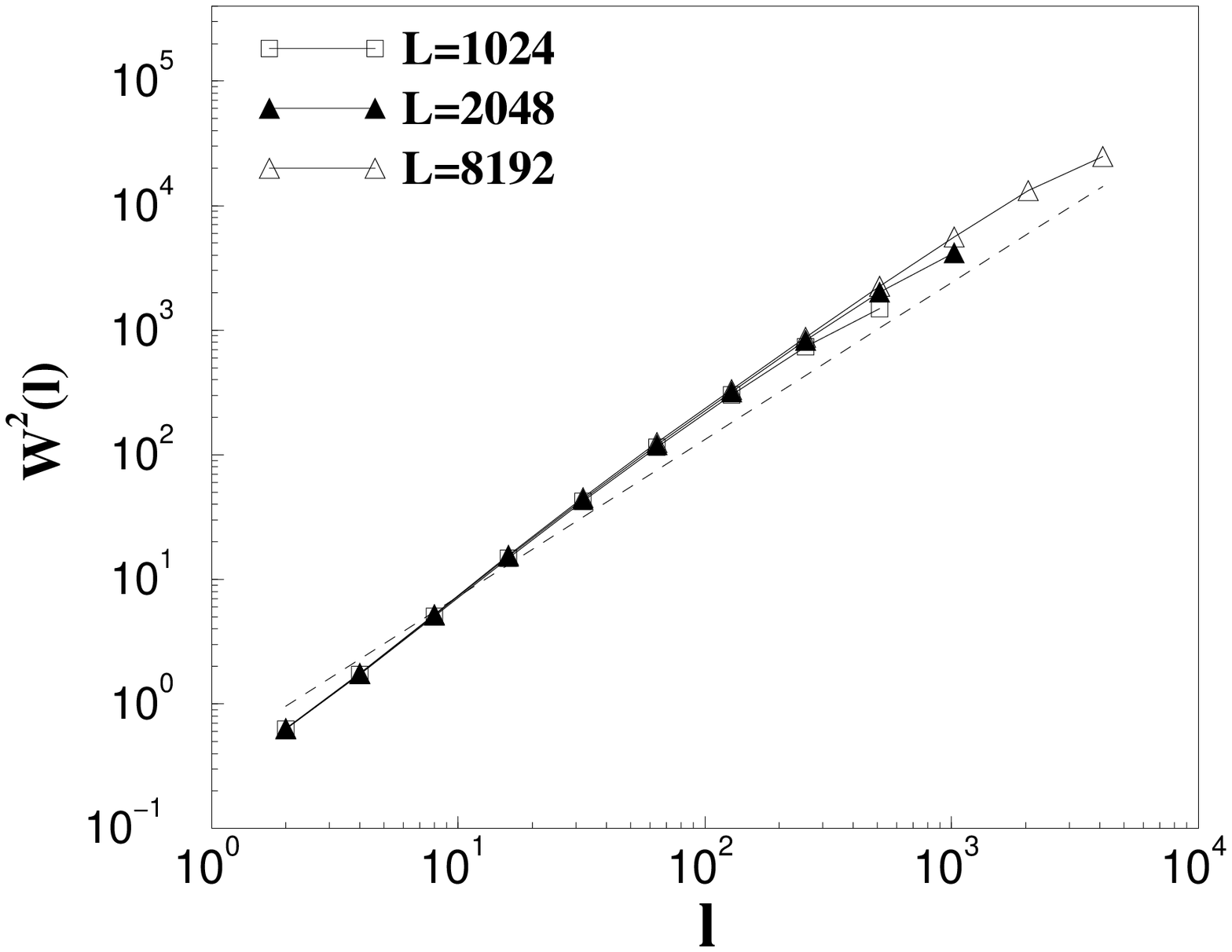,height=5.5cm,angle=0}}
\protect\caption{$W^2(l)$ vs. $l$  (both $W^2(l)$ and 
$l$ are expressed in adimensional units) 
for $p=0.02$ and different system sizes.}
\label{finite1}
\end{figure}

\begin{figure}[hb]
\centerline{
            \psfig{figure=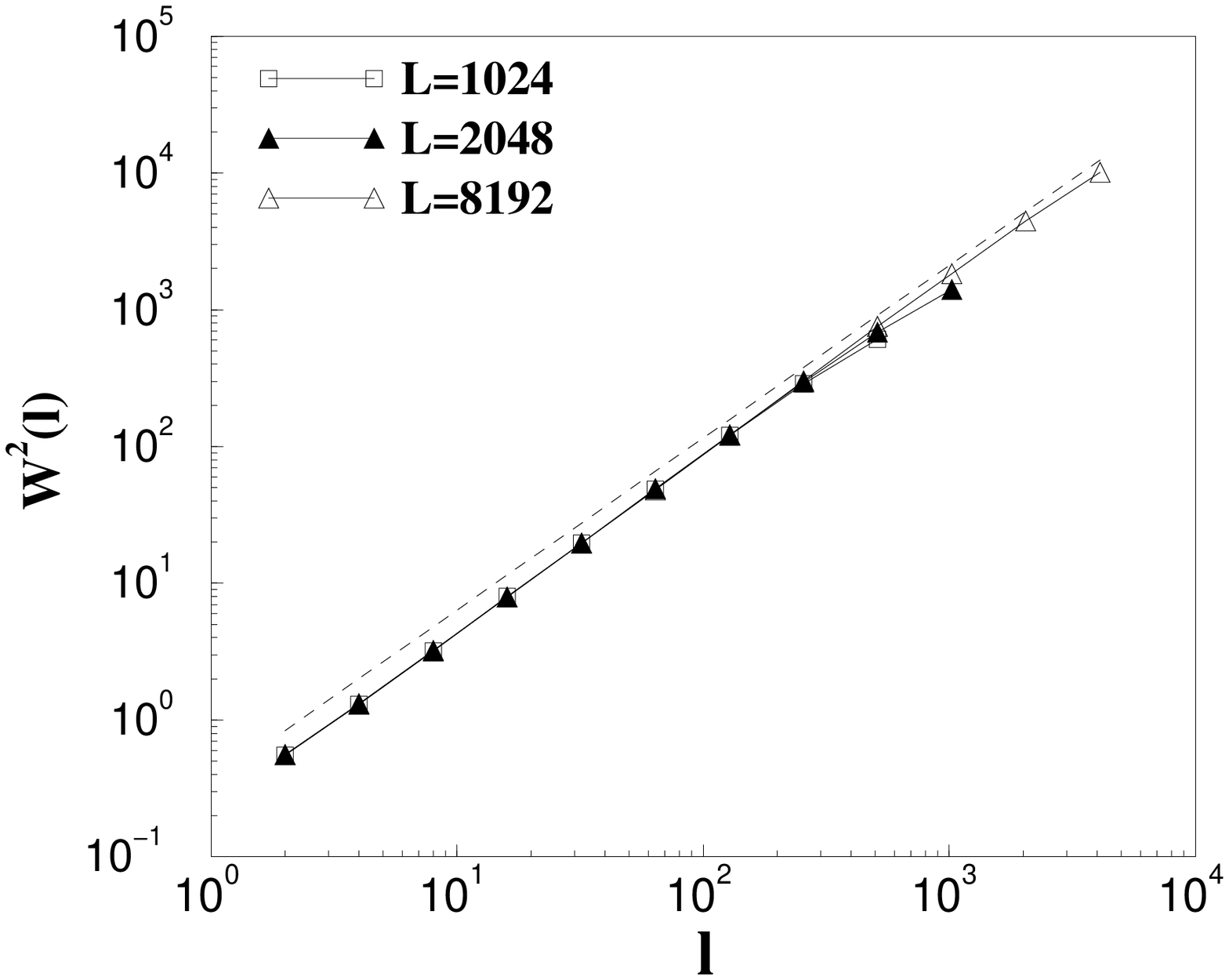,height=5.5cm,angle=0}}
\protect\caption{$W^2(l)$ vs. $l$  (both $W^2(l)$ and 
$l$ are expressed in adimensional units) 
for $p=0.2$ and different system sizes.}
\label{finite2}
\end{figure}

\begin{figure}[hb]
\centerline{
            \psfig{figure=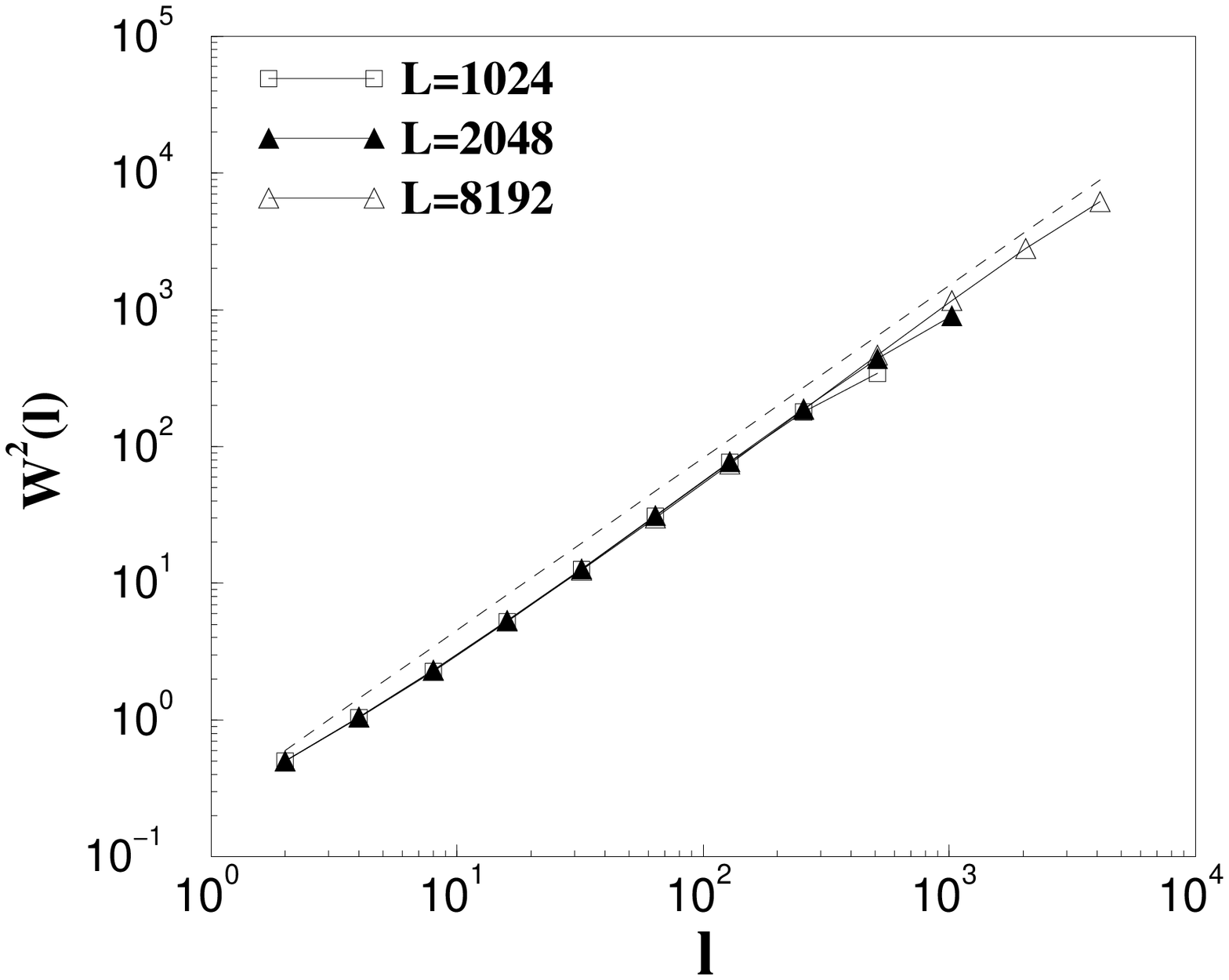,height=5.5cm,angle=0}}
\protect\caption{$W^2(l)$ vs. $l$  (both $W^2(l)$ and 
$l$ are expressed in adimensional units) 
for $p=0.5$ and different system sizes.}
\label{finite3}
\end{figure}

\begin{figure}[hb]
\centerline{
            \psfig{figure=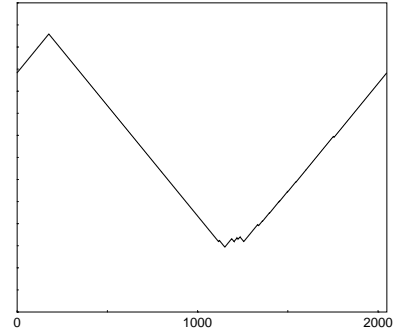,height=5.5cm,angle=0}}
\protect\caption{A realization of the growing surface for $p=0.0$ 
(horizontal adimensional position on the $x$ axis). 
The surface is composed by very big pyramids, thus with a strong 
prevalence of $S$ sites.
}
\label{piramid00}
\end{figure}

In Table \ref{table} we report the measured values for the dynamical 
exponents $\beta$ which turns out to be independent of $p$.
Fig.s (\ref{finite}-\ref{finite3}) show the scaling behavior
of $W^2(l)$ for different values of $p$.
For $p=0.0$ (i.e. maximal anisotropy) the measured values of 
$\chi$ are affected by a finite-size effect and they tend, in the
limit $L \rightarrow \infty$, to the value $\chi=1.0$ found for 
Sneppen model A \cite{snep} (Fig. \ref{finite}).
In this case the surface is composed by very big pyramids 
(Fig. \ref{piramid00}). 
On the other hand for $p=0.5$ one recovers the
universality class of Sneppen model $B$ with 
$\chi=0.63$ (Fig. \ref{finite3}). For all the other values of $p$ 
 between $0$ and $0.5$ ($p=0.02$ in Fig. \ref{finite1} and $p=0.2$ 
in Fig. \ref{finite2}), trying to perform fits away from the
saturation regions, one would be tempted to invoke the existence 
of a non-universal behavior ruled by the parameter $p$. 
A careful observation puts in evidence that the curves for 
$W^2(l)$ seem to exhibit a crossover between the Sneppen models
$A$ and $B$ universality classes. On the basis of $p$ one can define
a characteristic length $l^* \propto \frac{1/2-p}{p}$ above which
one could see the $\chi=0.63$ behavior and below which the
$\chi=1$ behavior. We shall come back on these considerations
later on when we shall discuss the power spectra.

\begin{figure}[hb]
\centerline{
            \psfig{figure=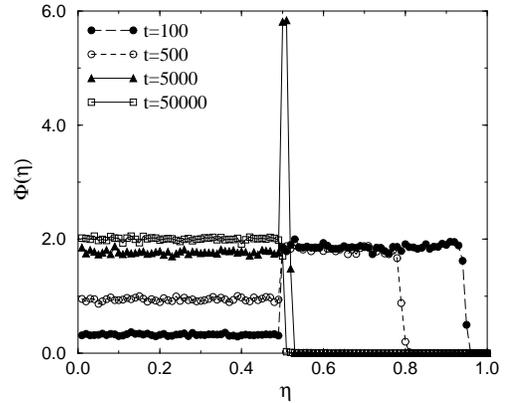,height=5.5cm,angle=0}}
\protect\caption{
Histogram $\Phi(\eta)$ ($\eta$ is an adimensional number) of quenched
variables, at different times $t$ (adimensional computer units), 
for $p=0.0$. Asymptotically, all (most of the) 
$M$ variables are eliminated.}
\label{histof1}
\end{figure}

\begin{figure}[hb]
\centerline{
            \psfig{figure=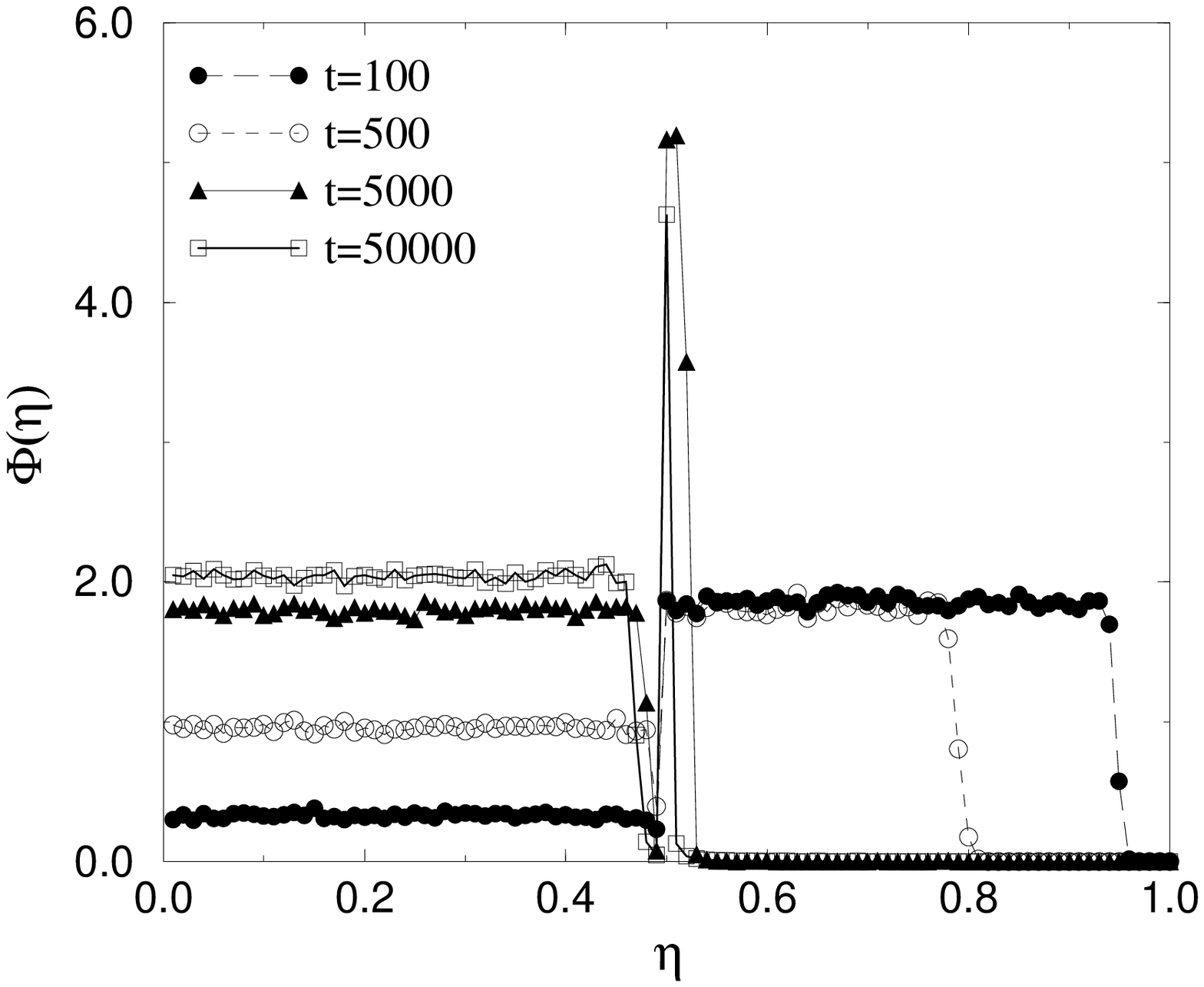,width=6.7cm,angle=0}}
\protect\caption{Histogram $\Phi(\eta)$ ($\eta$ is an adimensional number)
 of quenched variables, at different times $t$, for $p=0.02$.
}
\label{histof2}
\end{figure}

\begin{figure}[hb]
\centerline{
            \psfig{figure=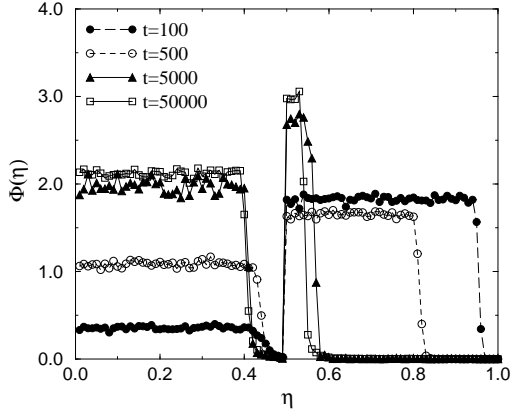,width=6.7cm,angle=0}}
\protect\caption{Histogram $\Phi(\eta)$ of quenched
variables, at different times $t$, for $p=0.2$.
}
\label{histof3}
\end{figure}

\begin{figure}[hb]
\centerline{
            \psfig{figure=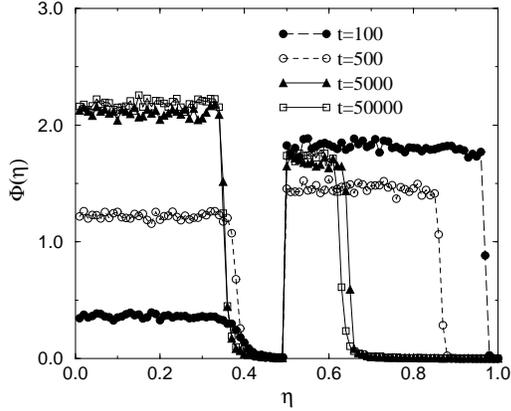,width=6.7cm,angle=0}}
\protect\caption{Histogram $\Phi(\eta)$ ($\eta$ is an adimensional number)
 of quenched variables, at different times $t$, for $p=0.5$.
}
\label{histof4}
\end{figure}

\begin{figure}[hb]
\centerline{
            \psfig{figure=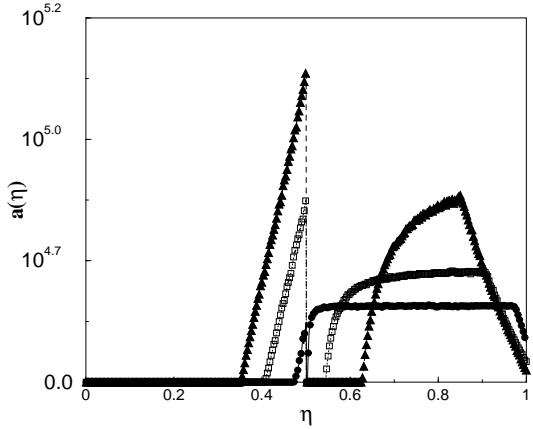,width=7.0cm,angle=0}}
\protect\caption{Asymptotic acceptation (not normalized) profile $a(\eta)$
 ($\eta$ is an adimensional number) for $p=0.02$ (circles), 
$p=0.2$ (squares) and $p=0.5$ (triangles).}
\label{acc1}
\end{figure}

We have also studied the time evolution of the distribution of 
random variables on the invading interface (the histogram 
$\Phi_t(\eta)$, where $\eta$ is a generic value for $\eta(x,h)$), 
which is of great importance for models with extremal dynamics. 
The results of the simulation are shown in Fig.s (\ref{histof1}-\ref{histof4}). 
One can see that $\Phi_t(\eta)$ self-organizes, for $p\neq0$, 
after about $10^5$ time steps, into a distribution 
that is the superposition of two theta functions, one for each class 
of variables, each one characterized by a critical threshold $p_c(p)$ 
depending on the parameter $p$. The meaning of these thresholds is that only
$S$ variables larger that $\eta_c^S$ and $M$ variables larger than $\eta_c^M$ can
 grow \cite{soc}. For $p=0$, instead, the histogram has no self-organized
critical state (Fig. \ref{histof1}). Looking carefully at Fig. (\ref{histof1})
we can see that, while in the initial transient there are a few $S$ sites, in
the asymptotic state most sites are $S$ sites. In fact nearly all variables
larger than $0.5$ (the $M$ variables) are disappeared. This observation agrees
with the actual structure of the surface, which is composed 
by very big pyramids (Fig. \ref{piramid00}), with a roughness exponent 
$\chi\simeq1$. This picture is confirmed by the acceptance 
profile $a(\eta)$, which is shown in Fig. \ref{acc1}. As in the 
Bak and Sneppen model, the acceptation profile (that is to say 
the distribution of the values of all updated quenched variables 
up to the actual time) exhibits correlation 
properties (it is not flat), reflecting temporal 
correlations in the dynamics. But, while the acceptation profile for 
 $S$ variables is quite similar to that of the BS model, going to zero 
linearly at $p_c$, the acceptation profile for $M$ variables has a
 more complicated behavior. This difference originates maybe from the fact 
that $S$ variables ($\eta\in[0.5,1]$) can turn into $M$ ($\eta\in[0,0.5]$) 
variables during the dynamics of the system, while $M$ variables 
cannot become $S$ variables. Moreover, $S$ variables can have 
developed correlations before the transition to $M$ variable and 
this affects the shape of $a(\eta)$ for $\eta \in [0.5,1]$. This
might account for the linear part of the acceptation profile of $M$ variables, 
 around $\eta=1$, but the non linear part is more puzzling. 
The coupling between $S$ variables
and $m$ variables could play a role in this behavior, too, but at the moment
we have no clear explanation of it.
>From the $a(\eta)$ we can get a good estimation of the critical 
thresholds $\eta_c^S$ and $\eta_c^M$ for different values of $p$ 
(see Table \ref{tab1}). 

\begin{figure}[hb]
\centerline{
            \psfig{figure=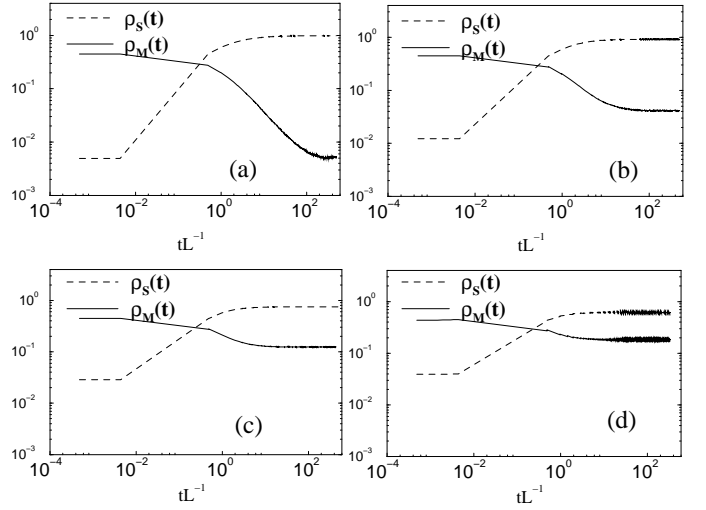,height=7.0cm,angle=0}}
\protect\caption{Time evolution ($t$ is expressed 
in adimensional computer time units) of the densities $\rho_S$ and $\rho_M$ 
of sites $S$ and $M$ respectively, for $p=0.0$ (a), $p=0.02$ (b), 
$p=0.2$ (c), and$p=0.5$ (d).
}
\label{rodens}
\end{figure}

The stationary state is characterized by a constant ratio between $S$ sites
and $M$ sites, that is to say the evolution equation for the densities
 $\rho_S$ and $\rho_M$ of sites $S$ and $M$ respectively, 
have an attractive fixed point in the stationary state (see Fig. 
 \ref{rodens}a-d), with the asymptotic values of $\rho_S, \rho_M$ depending on
the parameter $p$. One interesting observation is that, even in the case 
 $p=0$, that is to say only $M$ sites can be selected by the extremal 
dynamical rule, there is a stationary state for the system with $\rho_S\neq0$.
 This is due to the particular geometry of the lattice, for which the growth
of an $M$ site implies the creation or annihilation of some $S$ sites. In
other words, there cannot be surfaces without slopes ($S$ sites).

\begin{figure}[hb]
\centerline{
       \psfig{figure=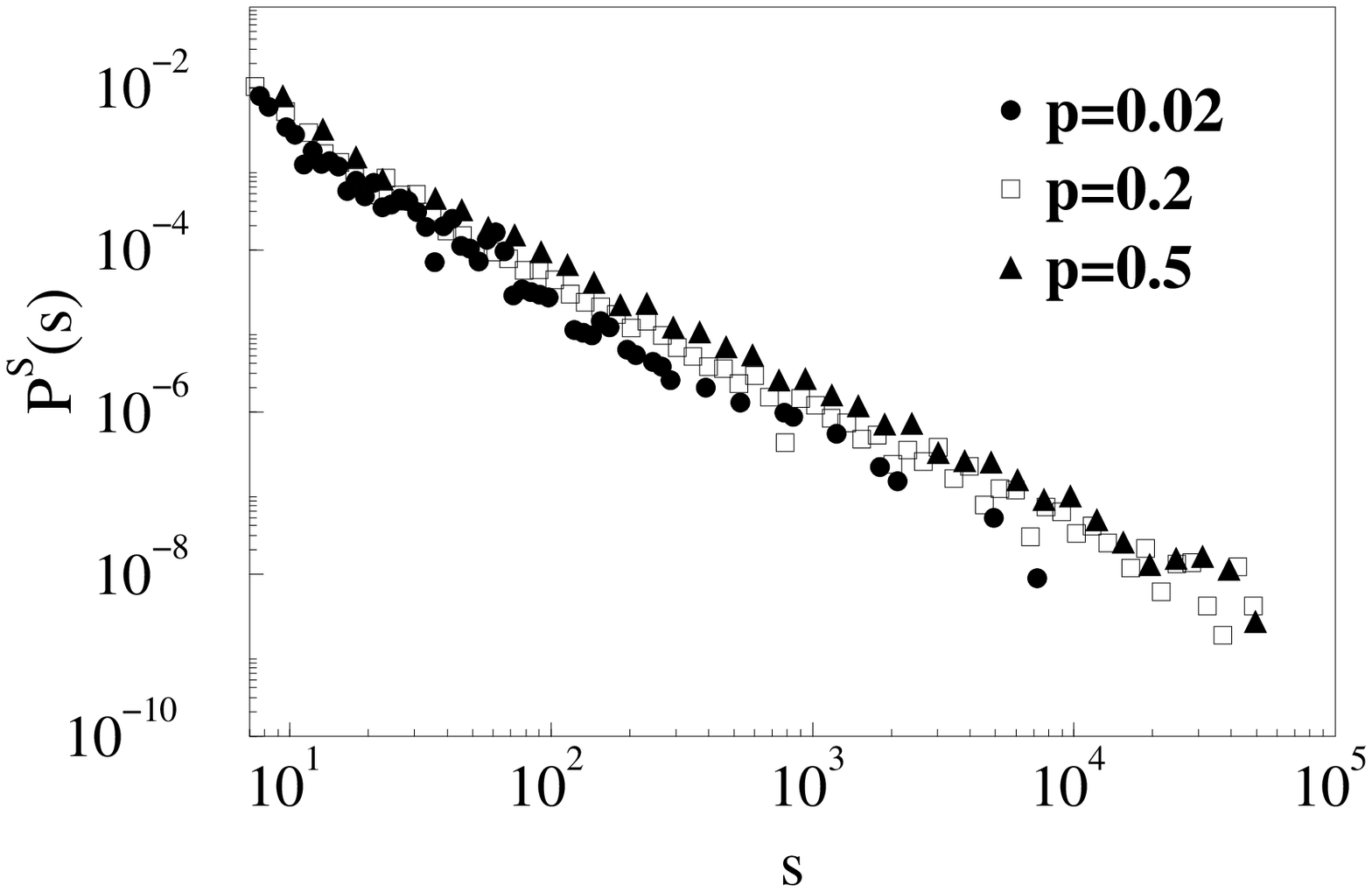,height=5.0cm,angle=0}}
\protect\caption{Binned S-avalanches distribution for different 
values of $p$.}
\label{ava1}
\end{figure}

\begin{figure}[hb]
\centerline{
       \psfig{figure=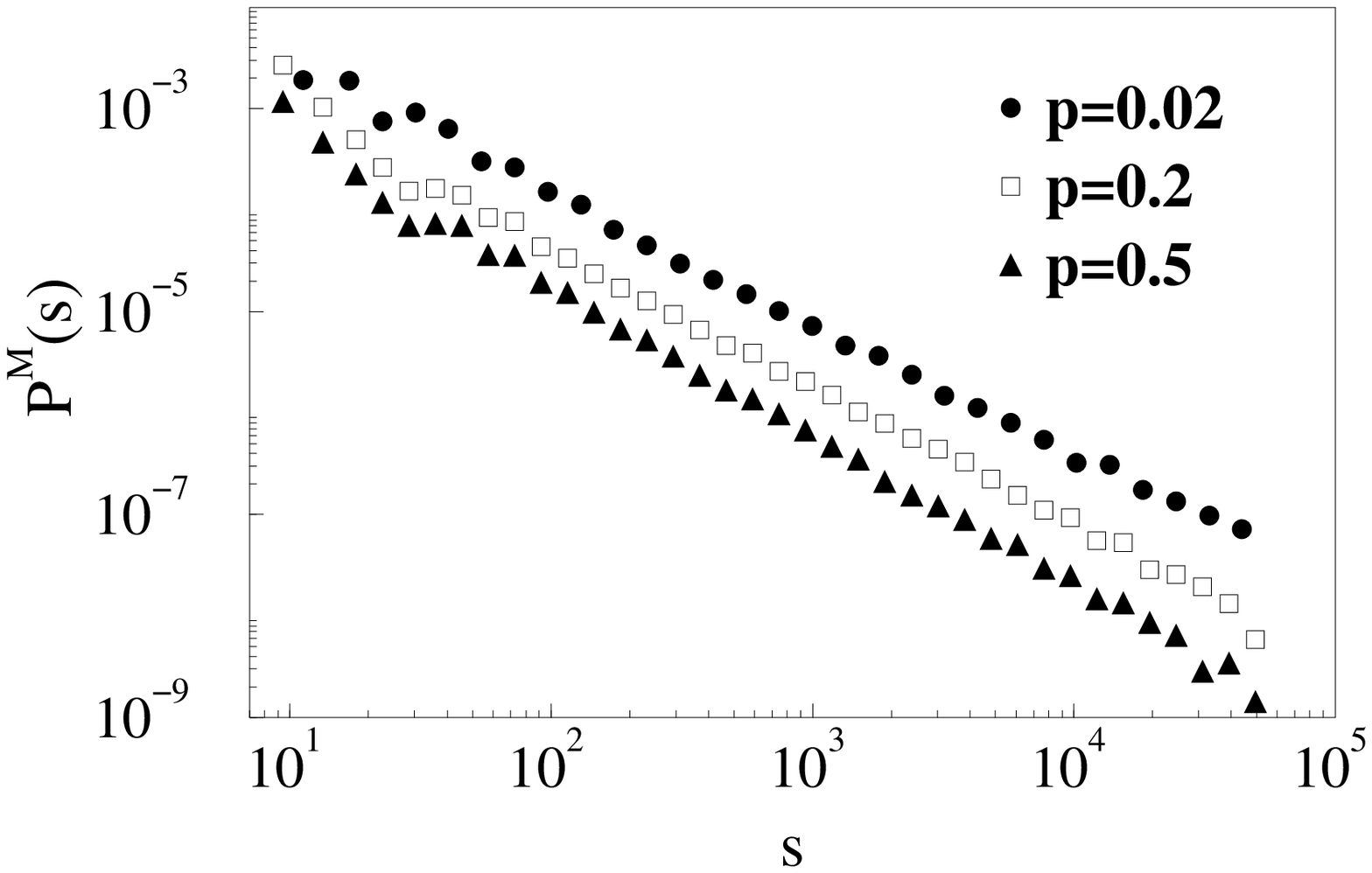,height=5.0cm,angle=0}}
\protect\caption{Binned M-avalanches distribution for different 
values of $p$.}
\label{ava2}
\end{figure}

The roughness exponent accounts for scale free spatial fluctuations in 
the interface profile. In order to characterize the eventual 
scale free fluctuations in the dynamical evolution of the system at its
asymptotic critical state, that is to say long 
range temporal correlations, we have studied the avalanche distribution. 
An avalanche is defined as a 
sequence of causally connected elementary growth events. For the class
 of models
with quenched disorder and an extremal dynamics to which  our 
model belongs, the initiator of a critical, scale invariant, avalanche is 
identified in the critical state by a site with quenched variable 
 $\eta_c^M(p)$ or $\eta_c^S(p)$ (respectively for an $M$ initiator and for an 
 $S$ initiator). The values of $\eta_c^M$ and $\eta_c^S$ for different values
of $p$ can be obtained 
by the asymptotic histogram distributions shown in Fig.s (\ref{histof1}-\ref{histof4}). 
In our case there are two classes of variables,
the $S$ and $M$ sites, and two possible initiators for an avalanche. We call
the avalanches that start with an $S$ site, $S$-avalanches, and the avalanches
that start with an $M$ site, $M$-avalanches.
An avalanche lasts when a variable which has been 
updated before the growth of the initiator is selected by the extremal 
dynamics. The statistics of off-critical avalanches has been 
shown to have the form \cite{soc,Masl}:
\begin{equation}
 P^X(s;\eta) = s^{-\tau_X}f_X(|\eta-\eta_c^X|s^{\sigma_X})
\label{disval}
\end{equation}
where $X=S,M$, and $\eta$ is the initiator of an $X$-avalanche. 
This distribution becomes a pure power law for 
$\eta=\eta_c^X$. 
 In the limit $t\rightarrow \infty$ the system 
 self-organizes into the critical state $\eta=\eta_c^X$, 
and the (normalized) avalanche size distribution becomes:
\begin{equation}
P^X(s;\eta_c^X)=\frac{s^{-\tau^X}}{\sum_{s=1}^{\infty}s^{-\tau^X}}
\label{1valPc}
\end{equation}
We have performed a set of about $10^3$ realizations of size $L=8192$, lasting
 each one $2\times10^6$ time steps, and collected the statistics 
of $S$ and $M$ avalanches over the last $10^6$ time steps, 
for $p=0.02,0.2,0.5$. These simulations required about $2$ months of CPU time
 on our computers (a network of DEC alpha machines with clocks going from
266MHz to 500MHz), and are at the best of our computation possibilities. 
To reduce numerical problems connected with the approximation on 
$\eta^M_c, \eta^S_c$, we used an alternative definition of critical 
avalanches in models with extremal dynamics, which resides on the causal 
relation between updated sites inside an avalanche 
(for details on the definition of critical avalanches 
see \cite{marsili94,gabrielli97,cafiero96}). The
results are shown in Figs. (\ref{ava1}), (\ref{ava2}). Even after this big 
computational effort, our numerical results are still a bit noisy. 
In particular the statistic of $S$ avalanches for $p=0.02$ is really poor. 
This is due to the fact that for small $p$ values most of sites 
selected by the dynamics are $M$ sites. Consequently, it is difficult to 
observe a quite clear power law behavior for both 
the $S$-avalanches and $M$-avalanches 
distributions. We point out that 
the presence of long range temporal correlations is not necessary 
for the model to have self-similar or self-affine 
spatial properties, as already observed 
in a different context \cite{mud}.

In order to better establish the critical properties of our model 
we have measured the power spectrum $S(k)$ of the equilibrium surface.
The model studied here is a discretized cellular automaton which can be 
thought as a modified version of the Sneppen model for quenched interface 
growth. The Sneppen model has been shown to be, at least in $1+1$ dimensions,
in the same universality class of the continuous Kardar Parisi Zhang 
equation with quenched noise (QKPZ)\cite{qkpz}. It is natural, 
but not necessarily true, to suppose that for our model, too, 
it is possible to find a formulation as a continuous growth equation. 
Given a general growth equation for h(x,t) like:
 \begin{equation}
\frac{ \partial h(x,t)}{\partial t}=A[h(x,t)]+\gamma(x,t)
\label{greq}
\end{equation}
where $A[...]$ is an operator acting on $h(x,t)$ and $\gamma(x,t)$ is an 
uncorrelated quenched noise (the ``temporal'' direction corresponds 
to the growth direction of the surface), 
if the operator $A[...]$ is linear and local, 
the equation can be Fourier transformed into
 \begin{equation}
i \omega \tilde{h}(k,\omega)=\tilde{A}(k) \tilde{h}(k,\omega)+
\tilde{\gamma}(k, \omega),
\label{greqft}
\end{equation}
and by introducing the propagator $G(k,\omega)$,
 \begin{equation}
\tilde{h}(k,\omega)=G(k,\omega)\tilde{\gamma}(k,\omega).
\label{greqft1}
\end{equation}
where $G(k,\omega)=[i \omega-\tilde{A}(k)]^{-1}$. 

The propagator $G(k,\omega)$ of eq. \ref{greq} 
is related to the power spectrum $S(k)$ of the interface 
in the asymptotic state. The power spectrum is so defined:
\begin{equation}
S(k)=\langle{FT[h(x)h(x')]}\rangle=
\langle{| \tilde{h} (k)|^2}\rangle
\label{power}
\end{equation}
where $FT[...]$ is the Fourier transform operator, 
the average is over different
realizations of the noise, $h(x)=h(x,t=\infty)$, $\tilde{h}(k)$ is the 
Fourier transform of $h(x)$. Eq. \ref{power} is valid is the case the noise is
uncorrelated in space and time, which is the case of our model. 
The relation between $\tilde{G}(k,\omega)$ and $S(k)$ is the following 
(\cite{zhang}):
\beq
\tilde{G}(k,\omega=0)^2=S(k)
\label{gsrel}
\eeq
Equation (\ref{gsrel}) tells us that the power spectrum of the interface can 
give informations on $k$-dependent part of the propagator $G(k,\omega=0)$ and 
consequently on the structure of the operator $\tilde{A}(k)$ in Eq. 
\ref{greqft}. For self-affine surfaces, the power spectrum 
follows a power law scaling
\beq
S(k)\sim k^{-2\delta}
\label{gsrel1}
\eeq
where $\delta$ is related to the {\em global} roughness exponent 
$\chi_{glob}$ through the scaling relation 
$2 \delta =2 \chi_{glob} +1$ \cite{zhang}.

\begin{figure}[hb]
\centerline{
            \psfig{figure=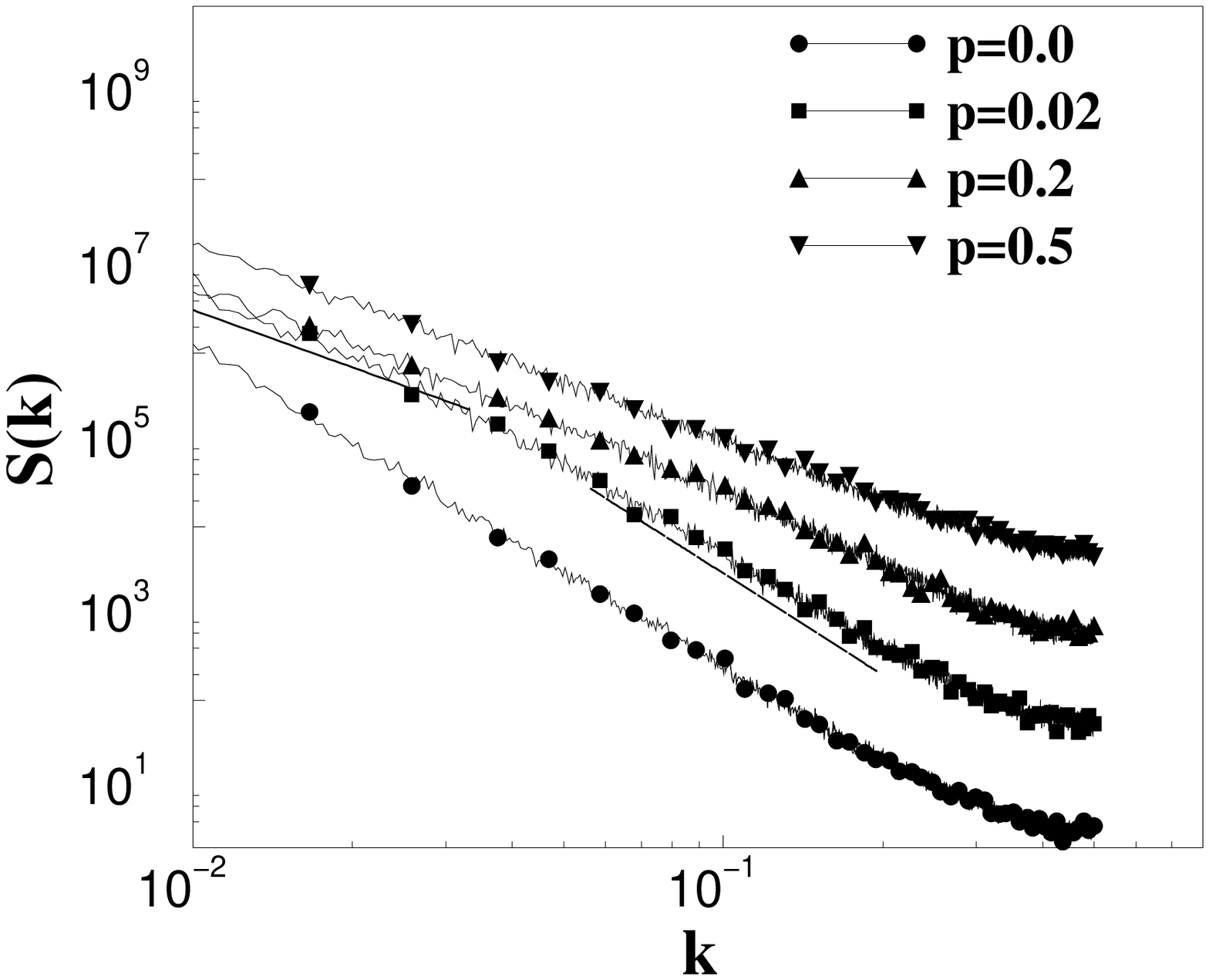,width=7.0cm,angle=0}}
\protect\caption{
Power spectrum $S(k)$ (all the quantities are expressed in 
adimensional units of the computer simulation) 
of our model for $p=0.0,0.02,0.2,0.5$ 
(values referring to, respectively, the plots
 from bottom to top) and $L=2048$. As a guide for the eye, 
we report the scaling law for KPZ (dotted line) and 
QEW (dashed line) universality classes.}
\label{pow1}
\end{figure}

\begin{figure}[hb]
\centerline{
            \psfig{figure=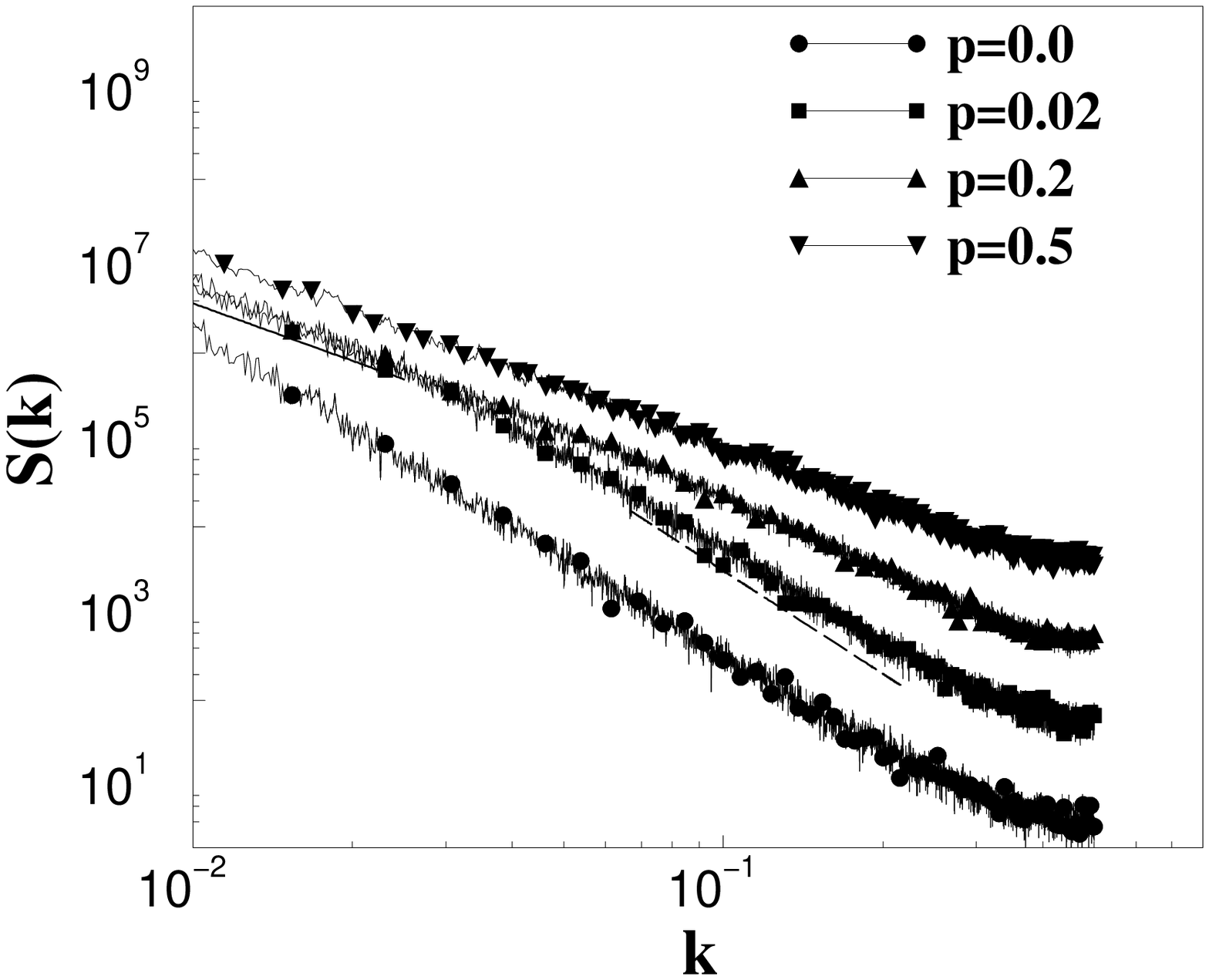,width=7.0cm,angle=0}}
\protect\caption{
Power spectrum $S(k)$ (all quantitities are expressed in 
adimensional units of the computer simulation) 
of our model for $p=0.0,0.02,0.2,0.5$ 
(values referring to, respectively, the plots 
from bottom to top) and $L=8192$. As a guide for the eye, 
we report the scaling law for KPZ (dotted line) and 
QEW (dashed line) universality classes.}
\label{pow2}
\end{figure}

Fig.s (\ref{pow1}-\ref{pow2}) report the behavior 
of the power spectrum $S(k)$  of the interface profile in the critical state 
for system sizes $L=2048,8192$ and different values of $p$. 

For large values of $k$ finite size effects connected to the 
discretized nature 
of the model become relevant and there is a deviation from 
the power law behavior. 
Away from this saturation effect it is evident in this case 
how $S(k)$ is characterized 
by a clear cross-over between two power law behaviors. 
In the low $k$ region ($k < k^* \propto p/(1/2-p) = 1/l^*$) $S(k)$ 
scales with an exponent $\delta_{low}$ close to $1$ (actually $1.07(3)$)
for all values of $p$. The QKPZ (Sneppen $B$ model) 
universality class is characterized by a global roughness 
exponent (coinciding with the local exponent 
we have measured through $W^2(l)$) 
$\chi_{glob}=\chi=0.63$, giving $\delta=1.13$, quite near to what we find. 
For intermediate values of $k$
($k > k^* \propto p/(1/2-p) = 1/l^*$) one gets  an exponent $\delta_{mid}$ 
quite near to the value $1.7$ (we find values between $1.7$ and $1.86$) 
which corresponds to the QEW universality class 
($\chi_{glob}\sim1.2$ \cite{zhang}), independently of 
the value of $p$. We point out that the QEW model is super rough 
with a global roughness exponent $\chi_{glob}\sim1.2$, and a local 
exponent $\chi=1.0$ (the one we measured through $W^2(l)$).
For $p$ near $0.5$ the intermediate $k$ region is very small, and it is 
difficult to distinguish it from the saturation region.
The same happens when one tries to fit the low $k$ region for 
$p$ close to zero.
If $p=0.0$ or $=0.5$, no cross-over effect is observed, 
since these two values of $p$ 
correspond to the ``pure'' QEW and KPZ 
universality classes, respectively.

From these results we have a confirmation that the model 
{\em does not exhibit non-universal critical properties}. 
The apparent non-universal roughness exponent is the consequence 
of a cross-over effect, tuned by the parameter $p$. 
The fact that this cross-over effect is difficult to observe 
when studying the scaling of the mean square width $W^2(l)$ 
of the sample, could explain the discrepancy between the experimental 
findings available in the literature.

\section{Conclusions}

In this paper we have introduced a model for surface roughening whose 
main peculiarity is that of taking explicitly into account the 
anisotropy of the growth process by means of a tunable phenomenological 
parameter $p$ which introduces local, i.e. dependent
on the local environment, dynamical rules in the growth. 
The simple introduction of just one anisotropy
parameter $p$ is far from being able to capture {\em all} the
characteristics of etching processes, and in general of 
surface roughening experiments. In etching experiments, for example, 
transport phenomena in the solution are likely to be important
and both concentration and agitation have strong effects on 
transport. Nevertheless, our model captures at least some basic elements of
the relationship between anisotropy and the apparent non-universality
observed experimentally in etching processes. 
Moreover, the general requirement 
of a microscopic dynamical
rule depending on the local environment could be a key element in the 
apparently observed non-universality in kinetic roughening phenomena. 

As a main outcome, the model exhibits a cross-over behavior 
in its critical properties. For each value of the anisotropy factor 
$p$ the system reaches a critical stationary state, with a characteristic 
length separating a KPZ-like (Sneppen B model) 
behavior from a QEW-like (Sneppen A model) behavior. 
The cross-over from one scaling behavior to the other is 
tuned by the anisotropy parameter $p$. If one looks at the 
scaling of $W^2(l)$, the cross-over effect cannot be easily discovered, 
and the system seems to have a non-universal roughness exponent. 
A careful analysis of the power spectrum $S(k)$, however, 
shows a clear cross-over effect. These results can probably help to explain
 the relevant discrepancies among experimental results 
\cite{burn,fluid,horvath}.
We believe that this behavior is the outcome of the complex
interplay between the global dynamics which selects at each time step
the weakest site and the anisotropy effect which takes into account
local constraints in the growth.

It is worthwhile to stress how our model suggests the possibility
of several analytical approach, from the treatment of the problem in 
terms of a continuous stochastic dynamical equation, to the 
single site mean-field approach \cite{dick}, or to the application of a method 
recently proposed for dynamical models driven by an extremal 
dynamics \cite{marsili94,gabrielli97,cafiero96,rts}. 
Particularly promising, in this respect, is a recently proposed 
non-perturbative Renormalization Group approach \cite{nprg}
which allows one to study self-affine problems.

The author acknowledges financial support under the 
European network project FMRXCT980183.

\begin{table}
\begin{center}
\begin{tabular}{|c|c|}
\hline
$p$ & $\beta$ \\
\hline\hline
$0.0$ & $0.96(2)$ \\
$0.02$ & $0.95(2)$ \\
$0.2$ & $0.94(2)$ \\
$0.5$  & $0.95(2)$ \\
\hline
\end{tabular}
\end{center}
\caption{Values of the dynamical exponent $\beta$ in our model for different 
values of the anisotropy parameter $p$.}
\label{table}
\end{table} 

\begin{table}
\begin{center}
\begin{tabular}{|c|c|c|}
\hline
$p$ & $\eta_c^S$ & $\eta_c^M$ \\
\hline\hline
 $0.02$ & $0.47(1)$ & $0.50(1)$ \\
 $0.2$ & $0.41(1)$ & $0.54(1)$ \\
 $0.5$ & $0.35(1)$ & $0.63(1)$ \\
\hline
\end{tabular}
\caption{Critical thresholds $\eta_c^S, \eta_c^M$ of variables 
$S$ and $M$, for different values of $p$.}
\end{center}
\label{tab1}
\end{table}

\end{multicols}
\end{document}